\documentclass[
 aip,
 amsmath,amssymb,
 reprint,%
]{revtex4-1}

\usepackage{graphicx}
\usepackage{dcolumn}
\usepackage{bm}
\usepackage{adjustbox}

\usepackage[utf8]{inputenc}
\usepackage[T1]{fontenc}
\usepackage{mathptmx}
\usepackage{etoolbox}

\makeatletter
\def\blfootnote{\xdef\@thefnmark{}\@footnotetext}
\makeatother

\makeatletter
\def\@email#1#2{%
 \endgroup
 \patchcmd{\titleblock@produce}
  {\frontmatter@RRAPformat}
  {\frontmatter@RRAPformat{\produce@RRAP{*#1\href{mailto:#2}{#2}}}\frontmatter@RRAPformat}
  {}{}
}%
\makeatother
\begin{document}
\preprint{AIP/123-QED}

\title{The Effect of Magnetization on Electron Heating in Low-Density Ultracold Neutral Plasmas}

\author{Ryan C. Baker}
\author{Bridget O'Mara}
\author{Jacob L. Roberts}
\email[Authors to whom correspondence should be addressed:
\href{mailto: Ryan.Baker@colostate.edu}{Ryan.Baker@colostate.edu} and
\href{Jacob.Roberts@colostate.edu}{Jacob.Roberts@colostate.edu}]{ }
\affiliation{Department of Physics, Colorado State University, Fort Collins, Colorado, 80523, USA}

\date{\today}

\begin{abstract}
Ultracold neutral plasmas provide a useful system for studying extreme parameter regime plasma physics in an accessible laboratory setting. The parameter space of plasma physics can be characterized in part by coupling strength and degree of magnetization. The range of achievable strong coupling is determined in part by the lowest possible temperatures that can be achieved. This work examines the early-lifetime electron heating of moderately coupled, strongly magnetized plasmas. This heating is dominated by disorder-induced heating and heating due to Rydberg atom formation. By using experimentally informed simulations, it is found that disorder-induced heating has a large influence in electron temperature well into the plasma lifetime. Additionally, the dependence of the minimum achievable electron temperature on magnetization and initial electron energy is examined. In this work, we find electron temperatures as low as $0.53\pm0.05\  \mathrm{K}$ (for electron density, $n_{e}$, of $6.1 \times 10^{12}\  \mathrm{m^{-3}}$), which determines the maximum coupling strength for the measured experimental conditions.
\end{abstract}

\maketitle

\section{Introduction} \label{sec:introduction}
 \blfootnote{This article may be downloaded for personal use only. Any other use requires prior permission of the author and AIP Publishing. This article appeared in \textit{Physics of Plasma} and may be found at https://doi.org/10.1063/5.0329398. Copyright 2026 Authors. This article is distributed under a Creative Commons Attribution-NonCommercial-NoDerivs 4.0 International (CC BY-NC-ND) License.}Ultracold neutral plasmas (UNPs) are useful tools for exploring extreme conditions of plasma physics in tabletop-scale laboratory settings, having been utilized since the late 1990's to better understand a variety of plasma physics phenomena \cite{UNPreview1,UNPreview2,UNP_context1}. Two important dimensionless quantities for specifying plasma parameters are magnetic field strength, $\beta$, and the electron coupling strength, $\Gamma$. While $\beta$ has more than one common definition in plasma physics \cite{intro_to_plasma}, for this work the definition used is that $\beta$ is the ratio of the cyclotron frequency of the electrons ($\omega_c$) to the plasma frequency of the electrons ($\omega_{pe}$). This describes how the magnetic field affects plasma processes such as transport and collisions\cite{Transport}. $\beta \sim \frac{B}{\sqrt{n_e}}$, where $B$ is the applied magnetic field and $n_e$ is the electron density of the plasma. $\Gamma$ is the ratio of the average nearest-neighbor potential energy of the electrons to the characteristic thermal energy of the electrons in the plasma\cite{Strong_Coupling}. This parameter describes the degree of spatial correlation between the electrons and also characterizes the degree to which many commonly used approximations are applicable (with lower $\Gamma$ indicating greater applicability)\cite{intro_to_plasma}. $\Gamma = \frac{e^2}{4\pi\epsilon_0a_sk_bT_e} \sim \frac{n_e^{1/3} }{T_e}$, where $a_s$ is the Wigner-Seitz radius and $T_e$ is the temperature of the electrons. For subjects of interest that scale with these parameters, different plasmas can be compared even if their densities and temperatures vary significantly \cite{Transport,TBR_electron_temperatures,Shielding_and_coupling,UNPs_and_High_density}. In this work, we use a combined experimental and simulation method to determine electron temperatures and hence electron $\Gamma$ at multiple values of $\beta$ in magnetized UNPs to explore the lowest electron temperatures that can be obtained.

The UNPs used in this work have comparatively low densities ($n_{e} =6.1 \times 10^{12} \ \mathrm{m^{-3}}$)\cite{UNPdensityexample1,UNPdensityexample2}. This fact combined with low achievable temperatures ($T_e \sim 1 \ \mathrm{K}$) allows for study of moderately coupled, strongly magnetized plasmas with straightforwardly attainable laboratory fields (140 G). The degrees of magnetization ($\beta = 17.7 $) and/or coupling ($\Gamma \sim 0.8$) make this plasma comparable to plasmas used in inertial confinement fusion \cite{ICF}, astrophysical plasmas \cite{Astrophysics1,Astrophysics2}, and non-neutral plasmas \cite{non-neutral} among others \cite{Stopping_power,Diffusion}. The primary goal of this work is exploring the parameter space with respect to strong coupling and magnetization that can be investigated in future experiments in these plasmas by studying the effect of magnetization on plasma heating. Work has been done in the field of UNPs to explore this space with respect to the ions\cite{Ion_UNP1,Ion_UNP2}. In this project, we explore the electron component of the plasma. Moreover, we investigate the formation physics of such low-density UNPs. The lower density of our UNPs allows for measurements earlier in the lifetime of these plasmas as compared to other measurements of electron temperatures\cite{Roberts2003,Gupta2007} whose faster electron dynamics and similar formation times prevent electron temperature measurements so early in the UNP lifetime.

There are two dominant heating mechanisms expected in the early lifetime of UNPs: heating due to three-body recombination and disorder-induced heating (DIH)\cite{DIH_temp_and_den_scale,Heating}. Just after plasma creation, DIH is expected to dominate in the very early lifetime of the plasma (1-2 $\omega^{-1}_{pe}$)\cite{Heating}. When the plasma is created, there is randomness in the electron-ion position distribution. DIH is the increase of kinetic energy in the system as the electrons move to a more ordered distribution in response to their Coulomb repulsion. After this period of DIH, three-body recombination is expected to become the dominant heating mechanism\cite{UNPexpansion,Rob_simexpansion}. Three-body recombination results from the collision of two free electrons near an ion such that one electron becomes bound to the ion while the other electron carries away kinetic energy from the first electron’s binding energy, ultimately resulting in heating of the plasma\cite{Recombination_Fluorescence,Heating}. Three-body recombination results in the creation of a Rydberg atom, which is a neutral atom with a valence electron in a highly excited state\cite{Rydberg_textbook}. While three-body recombination is the primary driver of Rydberg atom formation in weakly coupled plasmas ($\Gamma << 1$), as coupling strength increases there are multi-body interactions outside of isolated three-body recombination events which can produce Rydberg atoms. These multi-body interactions change the predicted density scaling of the Rydberg atom formation rate\cite{Scaling}. In this work we will refer to these various processes under the umbrella term Rydberg creation. Both DIH and Rydberg creation are predicted to be significantly reduced by increasing magnetic field strength due to electrons becoming increasingly bound to the magnetic field lines \cite{Heating}. In addition to these two heating mechanisms, heating from stray electric fields present at UNP creation can be significant\cite{electric_field_heating}.

UNPs are well-suited for studying the effect of these heating mechanisms in part due to their low densities, which result in smaller plasma frequencies ($\omega_{pe} = 2\pi\times20\ \mathrm{MHz}$) and thus more experimentally accessible timescales as compared to other plasmas. This allows for easier probing of the plasma early in its lifetime. In this work, early in the plasma's lifetime will be loosely defined as after the initial electron escape post plasma creation ($\sim 125\  \omega^{-1}_{pe}, \sim 1\  \mathrm{\mu s}$ under experimental conditions) until plasma expansion has a meaningful impact on the plasma electron temperature ($\sim 2500\  \omega^{-1}_{pe},  \sim 20\ \mathrm{\mu s}$ under experimental conditions).

The goals of this work are to investigate how Rydberg populations and early time electron temperatures shortly after UNP formation change with magnetization and initial electron ionization energies in the early lifetime of moderately coupled plasmas. DIH and Rydberg creation limit the lowest electron temperature in a given plasma, which in turn is a limiting factor of the highest achievable electron $\Gamma$. While these heating mechanisms have been studied experimentally in unmagnetized UNPs \cite{Recombination_Fluorescence,Electron_Distribution,Temperature_evolution,TBR_electron_temperatures}, this work focuses on weakly and extremely magnetized plasmas\cite{Transport}, specifically in the moderately coupled regime. In particular, we test whether substantial increases in electron magnetization can produce lower electron temperatures. To do this, Rydberg populations are directly measured experimentally (section II). The results of these experiments are used to modify and constrain a molecular dynamics simulation that is used to determine the electron temperature of the plasma (section III). By examining Rydberg populations, it is found that DIH has an important impact on the amount of heating during the plasma's early lifetime. We observed that large increases in the magnetic field strength decreased the electron temperature of the plasma. Lastly, the effect on electron temperature by utilizing initial excitation to Rydberg gases that then transition to UNPs\cite{spontaneous_evolution_Rydberg_plasma} was explored. The minimal electron temperature for these plasma conditions was measured to be as low as $530\pm50 \ \mathrm{mK}$  (section IV).
  
\section{Experimental Methods} \label{sec:Experimental Methods}
\begin{figure}[h]
    \centering
    \includegraphics[width=1.0\linewidth]{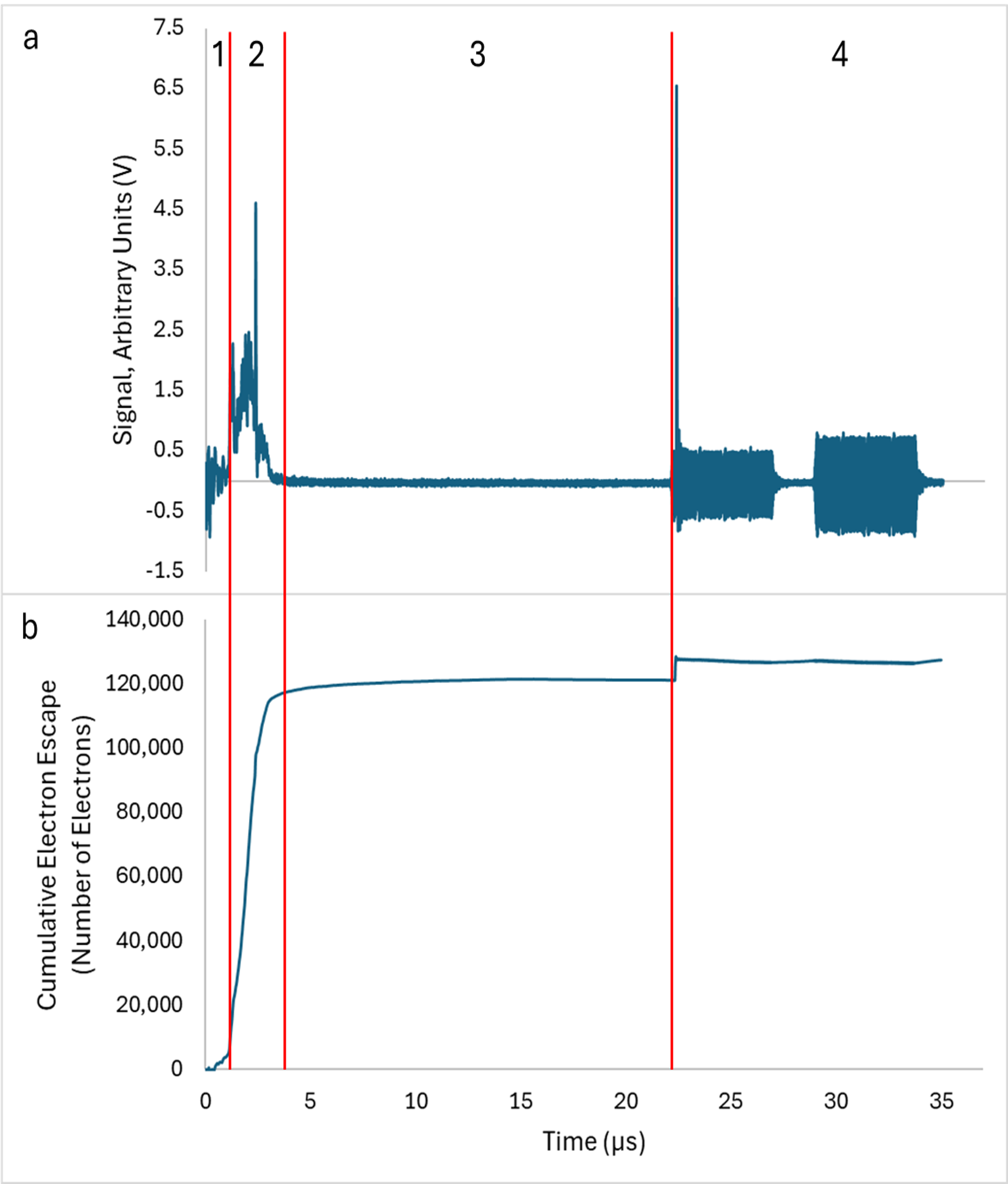}
    \caption{Example of MCP signal (a) showing electron escape over time and the integral of that signal (b). The signal is divided into four zones as shown in the figure and described in the main text. In zone 4, there are two RF pulses applied to the plasma, evident via pickup on the MCP signal in (a), which are smoothed out via the integration in (b)}
    \label{fig:timings}
\end{figure}

The experimental sequence starts by using laser cooling techniques to create a magneto-optical trap for $^{85}$Rb atoms \cite{Atom_Trapping}. 780 nm diode lasers excite the 5$S_{1/2}$ F=3 to 5$P_{3/2}$ F=4 transition for primary cooling. The atoms are then transferred to a purely magnetic anti-Helmholtz trap \cite{System}. This magnetic trap is translated via a mechanical track to the ionization region, where the UNP is created. First, the 5$S_{1/2}$ F=3 to 5$P_{3/2}$ F=4 transition is excited using 780 nm diode lasers. This is followed by an ionization pulse from a tunable pulsed dye-laser with a wavelength near 480 nm. By tuning the ionization pulse wavelength, we can vary initial electron energies. We specify electron energies through a Kelvin-based unit that gives the equivalent temperature in equilibrium when the ionization energy is positive ($T_0 = \frac{hc}{1.5k_b}\left(\frac{1}{\lambda_{dye}}-\frac{1}{\lambda_{thresh}}\right)$ with $\lambda_{dye}$ as the dye-laser wavelength and $\lambda_{thresh}$ being the ionization threshold wavelength). Initial electron energies range from 1 K below to 1 K above the ionization threshold. After creation there is a period of free evolution and then electron measurement. The apparatus is approximately cylindrically symmetric, and all applied fields are primarily along the axial direction. ``DC'' (slowly varying) electric fields are applied adiabatically via a wire grid near the plasma. Radio-frequency (RF) fields are applied via a copper disk within the apparatus. Magnetic fields are applied via a multi-turn wire loop outside the vacuum chamber. Electrons are extracted from the plasma with a DC field and directed towards a microchannel plate (MCP) detector. Electron collisions with the MCP create a proportional voltage signal. The MCP collects all electrons that escape the plasma given reasonable experimental conditions, both perpendicular and parallel to any applied fields. That signal is amplified, an example of which is shown in Fig. \ref{fig:timings}. This is referred to as the electron escape signal. For ease of analysis, the signal is integrated with respect to time (Fig. \ref{fig:timings}b). A more thorough overview of the experimental apparatus can be found in references \citenum{Lab} and \citenum{Lab2}. 

To determine the density, we use UNPs formed with initial energies above threshold ($T_0=3.5\ \mathrm{K}$). We apply RF pulses and find the RF frequency that resonantly drives electron center-of-mass motion\cite{Lab2}. That frequency is a function of UNP density. This provides a measurement of the spatial size of the UNP given the electron-to-MCP signal calibration and that in turn allows for a density determination based on the total electron number detected for all experimental conditions.

In Fig.\ref{fig:timings}, we have divided the experiment sequence into four zones to represent different stages of the experiment. The amount of electrons that are extracted during all zones is proportional to the density of the plasma. The 780 nm excitation laser intensity is adjusted to maintain a constant initial density on average. 

In zone 1 electrons are ionized or excited to Rydberg states\cite{spontaneous_evolution_Rydberg_plasma} and allowed to freely evolve for 1 $\mathrm{\mu s}$. The ionization pulse wavelength determines the initial kinetic energy (ionized) or bound state (Rydberg) of these electrons. In this experiment, it is advantageous due to signal size to set the ionization energy just below the ionization threshold. This results in a gas made of a type of Rydberg atoms we refer to as loosely bound atoms. These atoms can become ionized through either a small external electric field or collisions. This is distinguished from the broader category of Rydberg atom, which can have binding energies at or below the kinetic bottleneck \cite{Scaling}. After the photoionization pulse there is a rapid evolution in the plasma. Over a few plasma periods ($(2\pi\times\omega_{pe})^{-1}=50\ \mathrm{ns}$), interactions between particles will create a Maxwellian energy distribution in the electrons. In the case of loosely bound atoms, the electrons can still exchange energy via Coulomb force mediated ``collisions'' while bound. While some electrons are driven deeper into bound states, many more become ionized, which causes a cascading effect of ionization within the system. Over the range of experimental parameters measured, the fraction of loosely bound electrons at 1 $\mathrm{\mu s}$ varies from 20\% to 80\% of the total.

During zone 1 electrons will escape the neutral plasma, which will cause a deepening in the plasma's electron confinement until there is a significantly reduced electron escape \cite{Expansion_magnetic_field}. Additionally, there are stray electric fields due to patch charges. These stray fields are the dominant cause of electron escape during this time. These fields were found to be relatively constant on a many hours timescale, but drifted on a day-to-day timescale. Further, there were times of rapid change, which are presumed to be due to discharges during the applied RF (zone 4). Electrodes can be used to cancel out the axial component of the stray fields. However, with the current system there is no way to cancel radial stray fields. The stray fields were measured to be less than 1 V/m for most data sets but could drift higher. For the data used to determine electron temperature limits, the stray fields were determined to be 0.65 V/m or less. To mitigate the impact of stray fields, only data taken on the same day is generally compared directly, while data taken on different days are used to establish qualitative trends. During periods of rapid change, we avoided taking data.

In zone 2 a DC field, referred to as the extraction field, is applied to remove the majority of the electrons from the plasma. This field is applied adiabatically to avoid heating the plasma. The extraction field is determined to be 13.5 V/m. This field is set to be as high as possible without distorting the MCP signal so that the free UNP electrons are extracted as quickly as possible to stop the UNP evolution. The extraction field is present for the remainder of the experiment.

In zone 3, enough time is given for any free electrons to leave the plasma, which is expanding under the effect of a Coulomb explosion and as such the free electron plasma confining potential decreases with time. Additionally, free electrons on the opposite side of the plasma to the MCP can take a few microseconds to leave the plasma region. Loosely bound Rydberg atoms are ionized by this field, and their electrons escape as well.

In zone 4 an RF field is applied to ionize any Rydberg atoms that have been created. At this point the free electrons within the plasma have all left, and the remaining electrons should be bound within Rydberg atoms. RF pulses are used instead of DC fields to ionize the Rydberg atoms to ensure the best detection of the signal. In our apparatus, if a DC field from the wire grid is too large electrons can be accelerated into the wire grid instead of the detector. The use of the RF field avoids this problem as there is no net acceleration imparted to the electrons over a single RF cycle. A 40 MHz RF pulse is used to ionize the Rydberg atoms. This frequency was chosen because it resonantly couples to the electrodes around the UNP. The frequency of this RF pulse is much lower than the splitting between adjacent $n$ quantum levels for the most bound Rydberg states that can be ionized, but the pulse amplitude is above the Inglis-Teller limit\cite{InglisTellerlimit} and so there is a contribution to the signal from AC Rydberg ionization mechanisms\cite{Rydberg_lifetime}. The sharp response in Fig. \ref{fig:timings}b (at just before 23 $\mathrm{\mu s}$) to the ionization pulse is typical, showing that Rydberg ionization is caused in part by the leading part of the RF pulse, consistent with an effectively DC field ionization. After a brief delay, a second pulse is applied to ensure that any remaining Rydberg atoms, if present, are detected. These two pulses can be seen in the later times of Fig. \ref{fig:timings}a. The amplitude of these pulses was up to $546 \ \mathrm{V/m}$ (with a 30\% uncertainty in the RF field amplitude calibration). After the Rydberg atoms are ionized, the electrons are driven to the MCP by the extraction field.

The extracted signal in zone 4 is integrated to allow a determination of the fraction of ionized electrons that became bound in Rydberg atoms as a function of initial electron temperature and applied magnetic field (section IV). The Rydberg fraction is expected to be related to the electron temperature in multiple ways. The recombination rate is predicted to be temperature-dependent \cite{TBR_electron_temperatures,Scaling,Rydformation}. Rydberg creation produces electron heating, and so the electron temperature increases due to Rydberg atom formation. Once Rydberg atoms are formed, their states can be changed via collisions with free electrons, and that can increase or decrease the electron temperature depending on how the Rydberg atom state changes\cite{Scaling}. A method for directly measuring the electron temperature in a UNP has not yet been developed for this density. Instead, we determine the electron temperature via simulations whose parameters are tuned to reproduce measured experimental results.
\section{Simulation} \label{sec:Simulation}

A molecular dynamics (MD) code is used for the simulations. The details of the MD code can be found in references \citenum{Witte_Thesis} and \citenum{Lab2}. Electrons and ions are treated as discrete particles and are initialized as pairs with the location of the ion being chosen randomly as described below and the electron placed 500 nm away from the ion in a random direction. The electrons and ions interact via a modified Coulomb force, $F = \frac{1}{4\pi\epsilon_0}\frac{e^2 r}{(r^2+\alpha^2)^{3/2}}\hat{r}$, where $r$ is the distance between particles and $\alpha$ is a softening parameter usually used to avoid numerical instabilities\cite{Heating}. Starting electron velocities are oriented in a random direction with a magnitude determined by the dye-laser wavelength. The initial location of each ion is randomly determined using a spherical Gaussian spatial distribution such that the average ion density is proportional to $e^{(-r_{ion}^2/2\sigma^2})$ where $r_{ion}$ is the distance from the center of the UNP and $\sigma$ sets the spatial size of the UNP. The magnetic field was simulated using the Boris algorithm \cite{Boris}. The Leapfrog integration-like techniques used within that algorithm were implemented with a time step of $10^{-4}\ \omega_{pe}^{-1}$, which is small enough to achieve convergence with respect to total energy. However, it is not small enough to achieve convergence on individual particle motion such as position. Particle positions are used quantitatively only to establish the number of free and bound particles. While the specific electrons that are free/bound may change with step size, the number of free/bound particles did not. Neither did the binding energy distribution of the Rydberg atoms. This was tested by varying step size by a factor of 2. We note that we do not expect the position and velocity details of long-term correlations between individual particles to have an impact on our electron temperature determinations.

The MD is used in conjunction with the experiment. It has several parameters that are matched to the experimental measurements. The magnetic field of the MD is matched to the magnetic field of the experimental plasma. The MD density is matched to that of the experimental plasma, but the particle number is intentionally reduced to avoid the prohibitively long simulation times that a fully matched system would require. The MD results were examined as a function of particle number to confirm that there are no number dependent effects over a range of numbers (10,000 to 60,000 pairs, 30,000 in final results). The stray electric fields are simulated in the MD as a constant DC field and tuned such that the percentage of electrons that are extracted in zone 1 is matched in the MD and experiment. Any heating the stray field may be providing to the electrons is accounted for in the final temperature errors\cite{electric_field_heating}. The MD extraction field is matched to the experiment extraction field: 13.5 V/m. As with the experiment, the majority of the electrons escape from the plasma after the extraction field is applied. An electron is considered escaped if it is outside of a radius of 4$\sigma$ from the plasma's center. Zone 3 of the experiment is modeled by a 13 $\mathrm{\mu s}$ hold time. This is a shorter hold time than the experiment for the sake of simulation efficiency. Tests were conducted with respect to hold times and results converged from 13 $\mu$s to full experimental hold time. The RF pulse in zone 4 is not explicitly modeled. There is evidence that all loosely bound electrons were collected in the experiment (section IV), so by matching the extracted percentage of electrons in zone 2 the Rydberg population in zone 4 was also matched. 

\begin{figure}[h]
    \centering
    \includegraphics[width=1.0\linewidth]{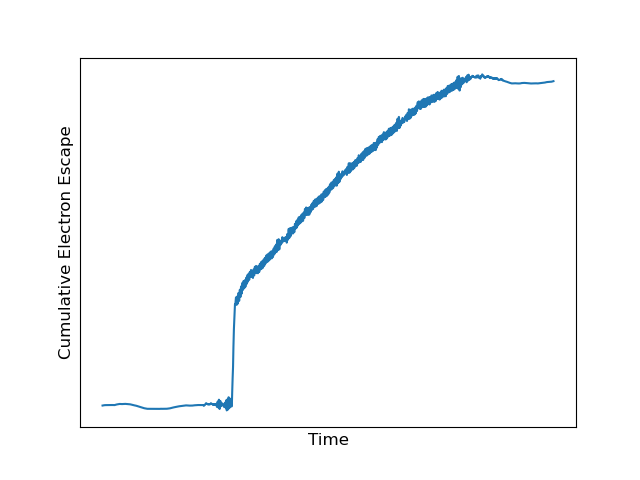}
    \caption{An example signal showing the Rydberg ionization in zone 4. The x-axis spans 0.1 $\mu s$. The integral of the MCP signal is shown in the figure. A sharp `spike'' in signal can be seen, corresponding to a DC-like ionization. Afterwards, a long tail of electrons ionized due to the AC field can be seen escaping. The ratio of electrons escaped during these two ionization regimes is used to determine binding energies, as described in the main text.}
    \label{fig:spike_ratio}
\end{figure}

The final parameter is $\alpha$. As the MD is not converged with respect to individual particle motion it may not be correctly quantifying the recombination rate of the Rydberg atoms, especially those more deeply bound, which have the highest susceptibility to error as they have the highest velocities. This recombination rate can be adjusted through $\alpha$ as a correction for imperfections of the simulation. $\alpha$ also adjusts the energy distribution of the Rydberg atoms, limiting the binding depth of the Rydberg atoms as $\alpha$ increases, as well as altering the percentage of extracted electrons predicted by the MD simulation. $\alpha$ was varied between 500 and 3000 nm in our simulations.  The $\alpha$ that was associated with the best correction factor for a given set of experimental measurements was determined through matching the total binding energy of all detected Rydberg atoms, which is the product of the average detected Rydberg atom binding energy and total detected Rydberg number. The values of $\alpha$ corresponding to this best correction factor varied from 2023 nm to 2904 nm. The experimental value of the average binding energy was determined through analysis of the time dependence of electron ionization signal after initial application of the RF pulse in zone 4, an example of which is shown in Fig. \ref{fig:spike_ratio}. In Fig. \ref{fig:spike_ratio}, one can see both an immediate ``spike'' in extracted electrons, which corresponds to a DC-like ionization, and a long tail of electrons being ionized afterward in an AC-like fashion (i.e. the electrons make a series of transitions from more deeply to more to more loosely bound states as the RF field oscillates, eventually becoming loosely bound enough to be ionized). Electrons with lower binding energies are extracted during the DC-like portion of the signal, while electrons extracted during the AC-like portion are more deeply bound\cite{Rydberg_textbook}. Tests were performed applying RF pulses of various amplitudes to low-density Rydberg gases (low enough density not to form UNPs) at several selected binding energies to determine the ionization depth vs. applied RF amplitude. By using these known depths, along with a model of binding energy distribution, it is possible to get an average binding energy from the experimental data. While the exact binding energy distribution is not known, we assumed a linearly decreasing population with binding energy to a binding energy limit with zero population. A flat binding energy distribution was also tested and included in the final error reported (section IV). 

\begin{figure}[h]
    \centering
    \includegraphics[width=1.0\linewidth]{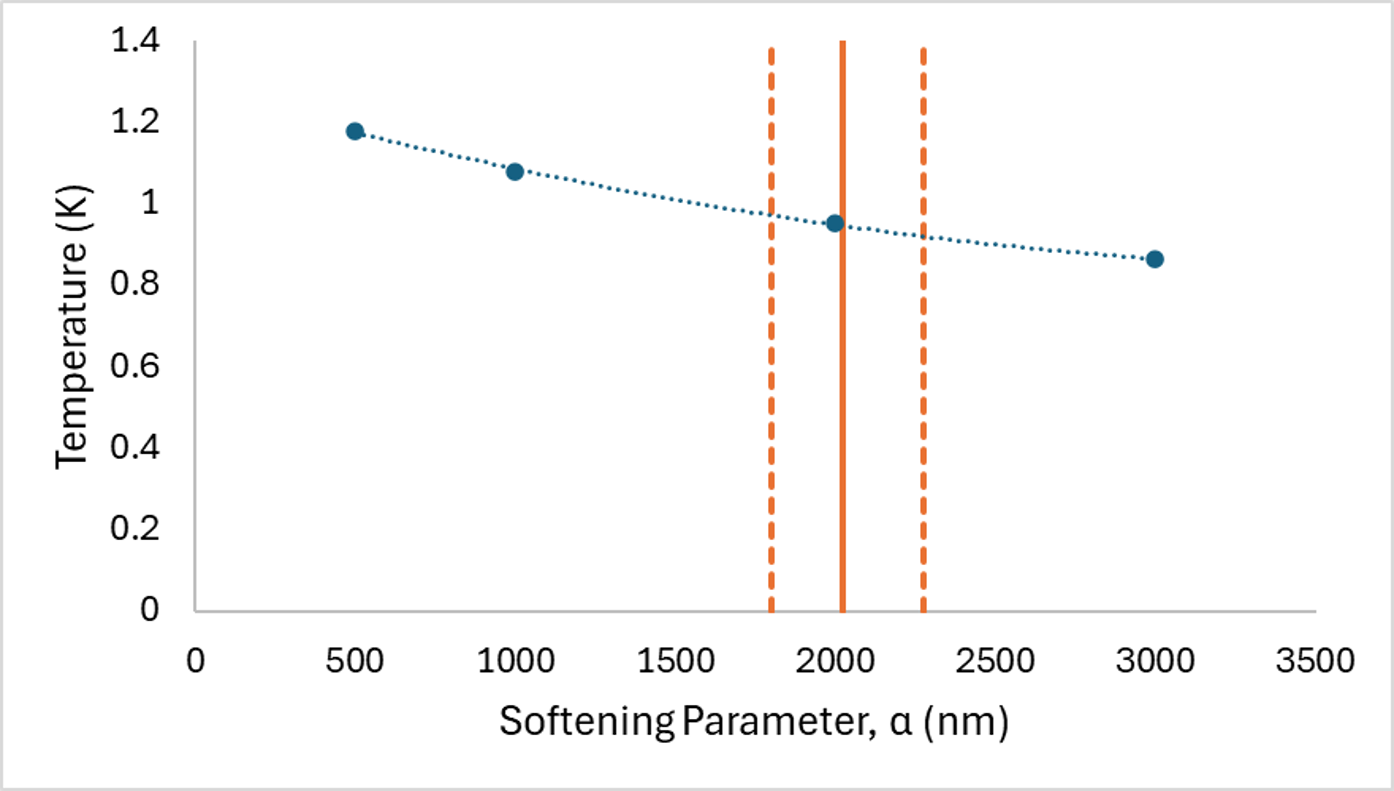}
    \caption{An example of matching the best fit alpha to a final temperature. The temperature at various alpha parameters is shown by the blue circle data points. A quadratic best fit line (blue dashed line) for the data is also shown. From this, a temperature for the best fit alpha (vertical orange line) can be found. The vertical dashed orange lines represent the total error in $\alpha$.}
    \label{fig:tempmatch}
\end{figure}

Once all these parameters are matched between the MD and experiment, the electron temperature in the plasma can be determined. The electron temperature is defined by the kinetic energy of free electrons at a time of 1 $\mathrm{\mu s}$ using the relationship $\frac{3}{2}Nk_BT_e = \sum_{i=1}^N\frac{1}{2} m_ev_i^2$, where $N$ is the total number of free electrons and $v_i$ is the velocity of a given free electron. Free electrons are defined as electrons that do not share the same nearest neighbor ion at both times 900 ns and 1 $\mathrm{\mu s}$ after UNP formation and have not escaped due to the stray field. Any electrons that are not free and have not escaped are considered bound. The electron temperature across a variety of parameters is calculated from multiple simulation runs to find the best $\alpha$ correction value. The electron temperature for that value is taken to be the electron temperature that is best associated with the particular set of experimental parameters that were measured. Electron temperatures for different experimental conditions are shown in Fig. \ref{fig:tempmatch}.

The need to use $\alpha$ as an adjustable parameter to match the experimental observations is a subject of continued investigation. While effects such as quantum mechanics, blackbody radiation \cite{Blackbody}, and bremsstrahlung radiation \cite{bremheat} are not included in the simulations, they are also not expected to be significant for our UNP conditions based on simple estimates. That being said, there are experiments elsewhere that have shown both blackbody radiation-induced state-mixing for times less than 1 $\mathrm{\mu s}$ \cite{Rydberglifetime2} and predictions that the Rydberg state angular momentum can affect electron ionization cross sections \cite{Electronimpact} and so there is a possible mechanism related to blackbody radiation leading to reduced Rydberg population. We are investigating this along with potential effects associated with the lack of convergence in electron velocity and position as future work beyond the scope of the results reported here.
\section{Results} \label{sec:ExResults}
There are three trends that were explored in addition to the overall electron temperature found through the simulation: Rydberg fraction vs. density, Rydberg fraction vs. magnetic field, and Rydberg fraction vs. initial electron temperature. These trends should be used to better understand the electron temperatures that were found using the MD simulation. 

\begin{figure}[ht]
    \centering
    \includegraphics[width=1.0\linewidth]{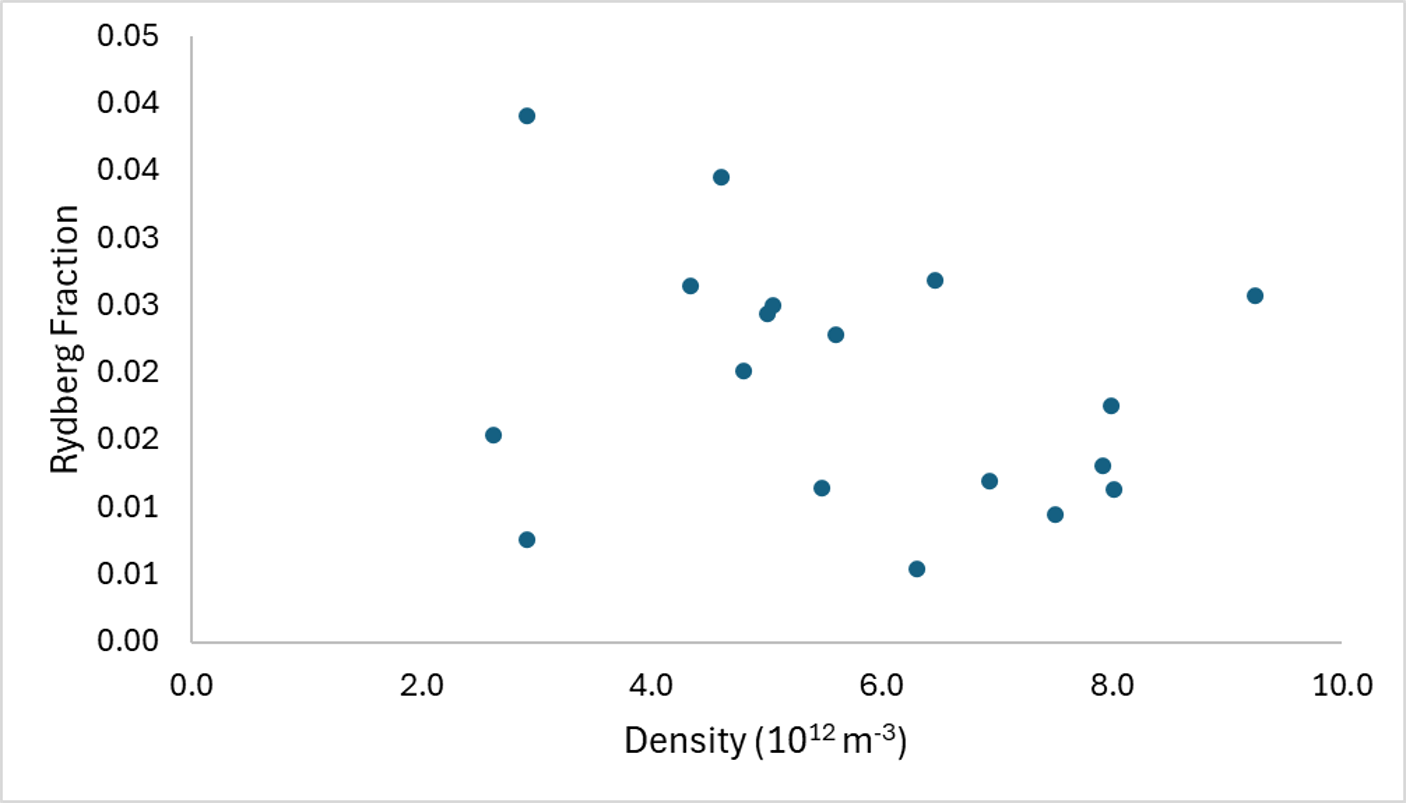}\\
    \caption{A typical data set showing shot-to-shot variation in Rydberg fraction and total electron number (and thus density). Variation in the total electron number stems from variation in the loading and ionization process of the atom cloud. Variation in the Rydberg fraction stems from variation in Rydberg number from shot to shot. MCP signal background noise contributes to variation in both Rydberg fraction and total electron number measurement. In the density range that was measured, throughout many such data sets, there was no evidence of $n_e^2$ scaling.}
    \label{fig:shotvar}
\end{figure}

Rydberg fraction versus density was analyzed using the ``natural'' electron number and thus density variation between shots (Fig. \ref{fig:shotvar}). Three-body recombination as the primary heating mechanism would predict\cite{Recombination_Fluorescence} a scaling in this quantity proportional to $n_e^2$. We found no strong scaling in Rydberg fraction with density as ascertained from multiple data sets like that shown in Fig. \ref{fig:shotvar}. It has been predicted and measured that as the coupling increases into the moderate and strong regimes the $\sim n_e^2$ scaling is not expected to hold \cite{Recombination_Fluorescence,Scaling}. The minimal scaling of Rydberg atom formation rate is expected to be $\sim \sqrt{n}$ \cite{Recombination_Fluorescence,Scaling} and the variation in our measured Rydberg fraction does not meet or exceed that minimal scaling. However, the measured Rydberg fraction is expected to be determined by multiple interconnected processes including three-body recombination, collisions between bound electrons and free electrons causing de-excitation, and ionization. All of these processes are dependent on electron temperature. The interplay between these processes, taking into account heating of the system, prevents a simple estimate for what the scaling of measured Rydberg atoms with density is expected to be. Additionally, any Rydberg atoms with a binding energy of less than $\sim-1\  \mathrm{K}$ will be ionized by the extraction field and will not be counted in the measured Rydberg fraction. This could be explored through reducing the DC extraction field, however that would allow for longer UNP evolution times, making experiment to MD simulation matching more difficult.

\begin{figure}[h]
    \centering
    \includegraphics[width=1.0\linewidth]{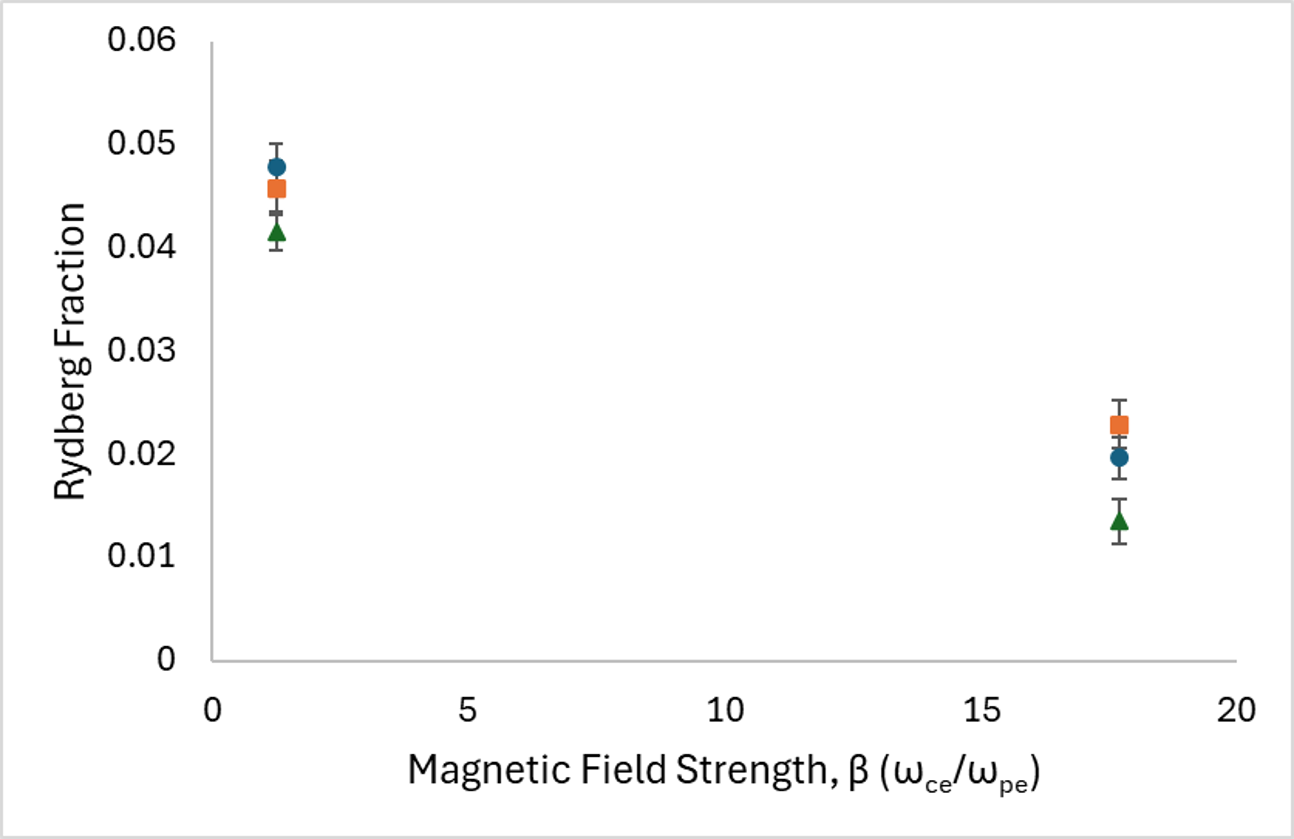}
    \caption{Data showing the Rydberg fraction as a function of magnetic field strength $\beta$ and applied RF field ($T_0=-1\ \mathrm{K}$). The 546 V/m (blue circles) and 410 V/m (orange squares) RF amplitude sets of data have no significant difference in Rydberg fraction in most cases. This is further discussed in the main text. The 228 V/m (green triangles) RF amplitude shows a lower Rydberg fraction. This lower Rydberg fraction is indicative of a field that is insufficient to ionize all of the Rydberg atoms.}
    \label{fig:RFvar}
\end{figure}

\begin{figure}[h]
    \centering
    \includegraphics[width=1.0\linewidth]{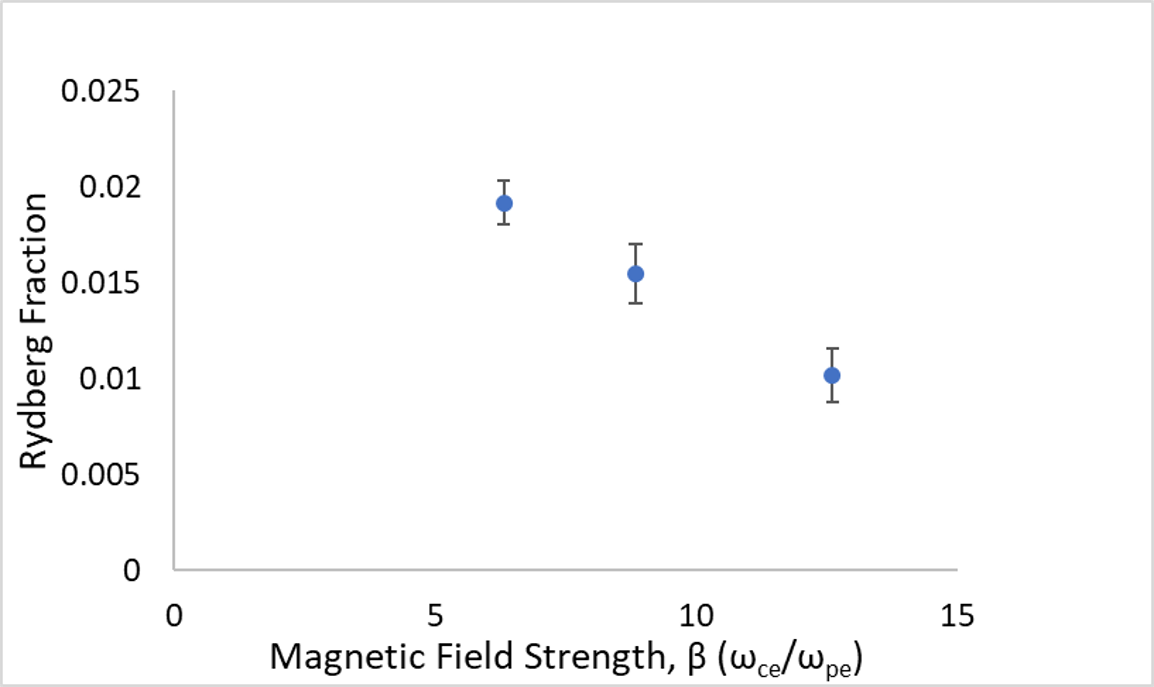}
    \caption{Rydberg fraction vs. $\beta$ for intermediate magnetic field strengths (546 V/m RF field, $T_0=-1\ \mathrm{K}$). The purpose of this data is to demonstrate that there is no resonant or abrupt behavior in the Rydberg fraction as the magnetic field is increased. This data cannot be quantitatively compared to Fig. \ref{fig:RFvar} (see main text for details).}
    \label{fig:Intfields}
\end{figure}

Figs. \ref{fig:RFvar} and \ref{fig:Intfields} show the relationship between Rydberg fraction and magnetic field. In these sets of experiments laboratory fields of 10 G ($\beta =$ 1.3) and 140 G ($\beta =$ 17.7) are compared. Fig. \ref{fig:RFvar} shows a clear decrease in Rydberg fraction between the low and high magnetization conditions. Additional intermediate fields were also checked to ensure that there is a smooth variation in Rydberg fraction as the magnetic field is increased, as shown in Fig. \ref{fig:Intfields}. Due to variations in stray fields as described above, the Rydberg fractions between figs. \ref{fig:RFvar} and \ref{fig:Intfields} cannot be quantitatively compared. The decrease in Rydberg fraction with magnetic field strength is predicted to occur \cite{Heating,Scaling}. Electrons in stronger magnetic fields have their directions of movement limited, as the cyclotron radius decreases and the electrons become more closely bound to the field lines, reducing the collisional Rydberg creation rate \cite{Heating,three_body_recombination_strong_field}.

We investigated the possibility that there were Rydberg atoms that were present that we failed to detect. In Fig. \ref{fig:RFvar}, it is shown that the measured Rydberg fraction is consistent with saturation between 410 V/m and 546 V/m RF pulses. This indicates that it is unlikely that we are missing a significant number of Rydberg atoms. This degree of saturation was present in all except the $\beta$ = 1.3 and $T_0= 0\ \mathrm{K}$ case. To evaluate how many Rydberg atoms might be going undetected in that case, the time dependence of the Rydberg ionization signal in zone 4 can be used to get information about the binding energy distribution of the Rydberg atoms since less bound atoms are ionized more rapidly. To do so, ionization signals vs. binding energy in Rydberg gases with low enough density that UNPs do not form were measured experimentally. We found that there was sensitivity to Rydberg atoms of binding energy -30 K for the 546 V/m signal. From estimates based on the time dependence of the measured ionization signals, it is unlikely that more than 20\% of the Rydberg atoms went undetected. With respect to other considerations, the radiative lifetimes of the Rydberg atoms being measured are calculated to be hundreds of microseconds \cite{Rydberg_lifetime}. Measurement of Rydberg atoms as a function of the hold time in zone 3 did not indicate any loss rates in Rydberg number that would noticeably lower the measured fraction. For the $T_0 = 0\ \mathrm{K,}\  \beta = 1.3$ condition, making comparisons to magnetic-field-free predictions of Rydberg atom formation rates seems reasonable \cite{Heating}. In light of this, we note that there is consistency within 10\% between our measured Rydberg fractions with predictions from Ref.\citenum{Scaling}, using the electron temperature determined by our matched MD simulation.

\begin{figure}[h]
     \centering
     \includegraphics[width=1.0\linewidth]{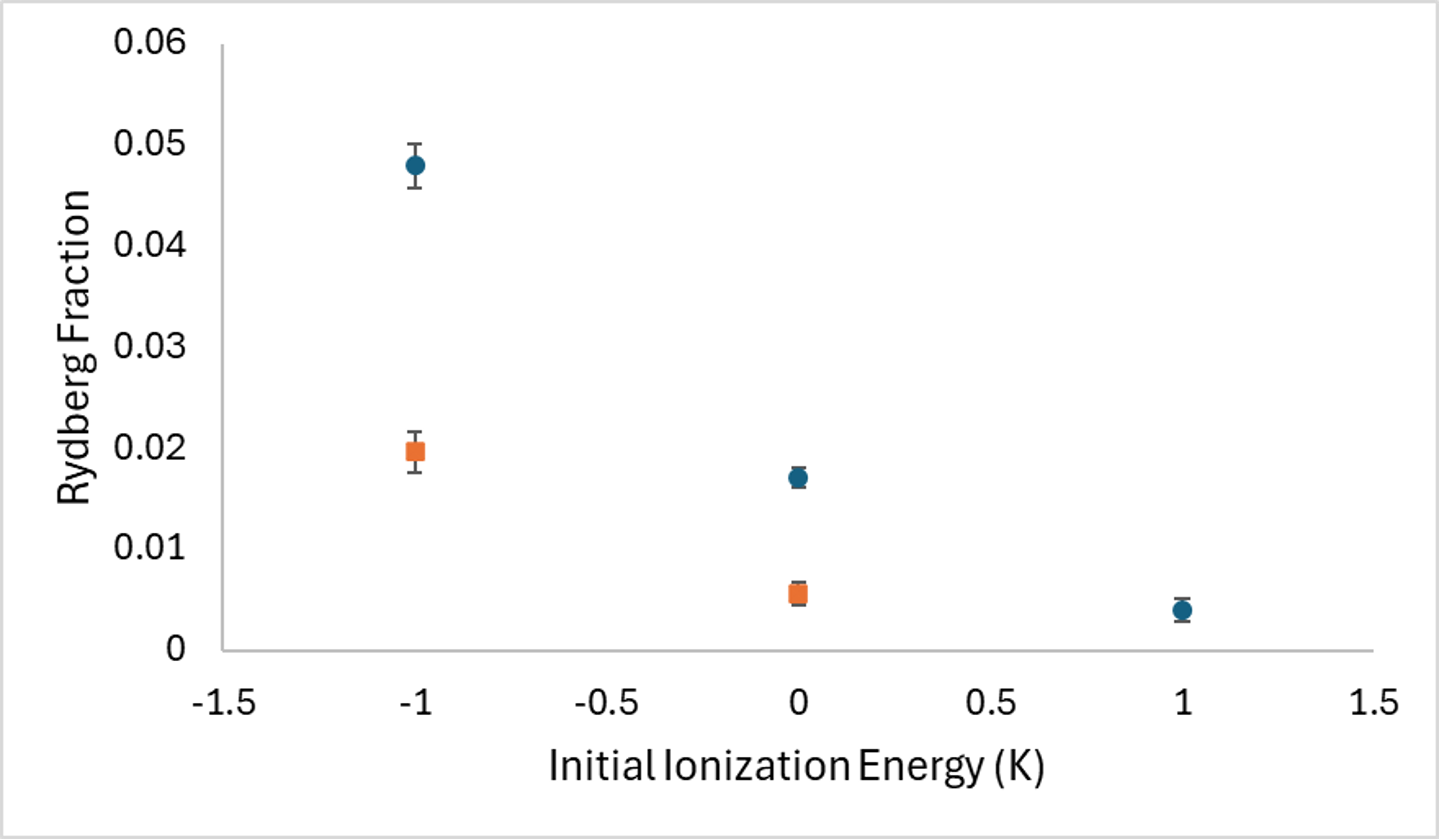}
     \caption{Rydberg fraction vs. initial ionization energy expressed in units of equivalent temperature (see main text for details) using a 546 V/m RF field. Low magnetic field strength ($\beta = 1.3$) is represented by the blue circles. High magnetic field strength ($\beta = 17.7$) is represented by the orange squares. There is no high field point for the 1 K condition. As the temperature increases, the Rydberg fraction decreases. There is approximately a factor of 2.5 decrease in Rydberg fraction between the low and high fields in both the -1 K and 0 K cases.}
     \label{fig:Tempvar}
 \end{figure}

Lastly, Fig. \ref{fig:Tempvar} shows the relationship between Rydberg fraction and initial ionization energy. As expected, the Rydberg fraction decreases both with increasing $T_0$ and increasing magnetic field. Signal-to-noise considerations make measurements of the Rydberg fraction at higher values of $T_0$ difficult.

\begin{table}[h]
\renewcommand{\arraystretch}{1.5}
\begin{adjustbox}{width=.8\columnwidth}
\begin{tabular}{|c|c|c|c|} 
\hline
 $\beta$ & $T_0$ (K) & $T_e$ (K) & $\Gamma$\\ 
 \hline
1.3 & 0.0 & $1.02\pm0.04$ & $0.48\pm0.02$\\ 
 \hline
 17.7 & 0.0 & $0.58\pm0.08$ & $0.85\pm0.14$\\ 
 \hline
 1.3 & -1.0 & $0.53\pm0.05$ & $0.94\pm0.09$\\ 
 \hline
  17.7 & -1.0 & $0.71\pm0.03$ & $0.69\pm0.03$\\
 \hline

\end{tabular}
\end{adjustbox}
\caption{Table showing electron temperature and $\Gamma$ at the end of zone 1 under different magnetization and initial ionization energy combinations.}
\label{Table 1}
\end{table} 

These observed trends contribute to a better understanding of the electron temperature results presented in table 1. Table 1 shows the electron temperatures and associated $\Gamma$ early in the UNP lifetime for different combinations of initial ionization energy and magnetic field strength. The uncertainties shown in the electron temperatures are the combination of experimental uncertainties and those associated with the variance in the parameters chosen and tuned in the MD. Uncertainty from the experiment represents on average 53\% of the total uncertainty. This includes the uncertainty in determining the extracted electrons due to the stray field (zone 1) and extraction field (zone 2 and 3), the uncertainty in measured Rydberg fraction (zone 4), and the uncertainty in the determination of the precise time dependence of the RF ionization signals in zone 4 (Discussed in section III). The dominant uncertainty from experimental uncertainties was the determination of the stray field. Uncertainties from the MD and modeling represent on average 47\% of the total uncertainty and include the uncertainty in relating $\alpha$ to electron temperature, randomness in the initialization of the MD simulations, as well as sensitivity in the final matched $\alpha$ to the selected extraction field and binding energy distribution of the Rydberg atoms. The dominant source of error was in the determination of the binding energy distribution of the Rydberg population.

Beginning with the $T_0 = 0\ \mathrm{K}$ conditions, we observe that there is an increase in $\Gamma$ with an increase in magnetic field strength, as predicted \cite{Heating}. Additionally, referring to Fig. \ref{fig:Tempvar}, one sees that the measured Rydberg fraction fell by just over a factor of 2 between the low and high magnetic field strengths. However, this relatively small population of Rydberg atoms is not enough to account for total electron heating in the early plasma lifetime. We note that there is a population of Rydberg atoms that would be ionized by the extraction field. These electrons would be extracted during zone 2 and not included in the Rydberg population ionized by the RF pulses. Totaling all Rydberg creation heating (heating from Rydberg atoms ionized by the RF plus those ionized by the extraction field) plus the heating from any stray electric field results in an estimated amount that is less than 1/3 of the total heating derived from the experiment-matched MD simulation. This indicates that DIH represents the most significant heating contribution in the plasma's early lifetime for these conditions. Additionally, the reduction in electron temperature with increasing magnetic field strength indicates that heating from DIH is sensitive to the magnetic field strength.

In the $T_0 = -1\ \mathrm{K}$ conditions, $\Gamma$ is seen to decrease with rising magnetic field strength. When interpreting this result, it is best to remember that when tuning the photoionization laser wavelength to be under threshold, especially with large magnetic fields, the majority of electrons are bound in Rydberg atoms. As high as $\sim$ 80\% under our conditions. This combined Rydberg gas/UNP is a complicated system, and results from it should not be applied to a fully ionized plasma such as the T = 0 K data. Even so, it remains true that the overall heating did increase. This is despite the reduction in measured Rydberg fraction seen in Fig. \ref{fig:Tempvar}. Overall, this supports the conclusion that the reduction in Rydberg population due to the presence of a magnetic field is not the sole or even primary determination of the amount of electron heating of the plasma, further illustrating the importance of DIH as a plasma heating mechanism at early times. 

We have measured that there is a statistically significant reduction in Rydberg fraction with increasing magnetic field strength and with increasing temperature. We find that DIH is a dominant heating mechanism under these conditions, and that the electron temperature can be reduced by increasing magnetic field strength. Lastly, we have shown that by using an initial ionization energy just under threshold, it is possible to push $\Gamma$ to a higher value than what can be achieved starting with a fully ionized plasma under certain, but not all, conditions. However, the resulting mixture of free and loosely bound Rydberg atoms (where the latter can outnumber the former) presents complications for using such UNPs in other experiments. Similar levels of temperature reduction can be achieved by increasing magnetic field strength without the introduction of Rydberg gases. For the measured conditions, the upper limit on $\Gamma$ was found to be $0.94\pm0.09$, with a corresponding electron temperature of $530 \pm 50\ \mathrm{mK}$.
\section{Conclusions} \label{sec:Conclusion}
We describe a method for determining the early time electron temperature in an ultracold neutral plasma through a combined experimental measurement and simulation technique. This is used to study how early lifetime heating in a plasma depends on magnetization and initial temperature. We observe that there is a significant dependence of the Rydberg atom number on both magnetic field strength and initial ionization energy, with a rapid decrease in number with both increasing ionization energy and increasing magnetic field strength. We observe dependence of the electron temperature on the magnetic field strength as well. Rydberg creation heating was not found to be the most significant source of heating, indicating the importance of disorder-induced heating for the early lifetime of ultracold neutral plasmas. Finally, we see that excitation to Rydberg gases can be used to achieve lower electron temperatures for these plasmas under certain conditions, although the underlying physics for these systems are complex.

Expanding this work further would benefit greatly from increasing the stability in the data collection procedure. To support such an effort, we plan to install electrodes to cancel the radial stray field in the system. The data in this article that is reported and used for the main analysis is a subset of the total data taken, specifically data taken on days when minimal stray fields were present. Even so, the associated systematic uncertainty owing to the stray field (6\%) for such data has a noticeable impact on the error in our electron temperature determination. If we were able to compensate for stray fields both axially and radially, data collected over longer time periods would be directly comparable, reducing uncertainty in measurements and allowing for more rapid and efficient studies of the parameter space. Additionally, we are investigating ways to improve our MD code to allow for smaller step sizes with reasonable run times, which has the potential to help address convergence-related imperfections. These fixes would also allow expansion of parameter space explored. Two directions of such an expansion are to take more data at the intermediate magnetic fields, which may allow for determining a functional form of electron temperature variation with magnetic field, and going to even lower initial electron energy to search for the upper bounds of $\Gamma$ for our experimental conditions.
\section*{Acknowledgements} \label{sec:acknowledgements}
We are thankful for Perry Hurd's contributions to the analysis of the binding energy distribution. This work was funded by the Air Force Office of Scientific Research (AFOSR), grant number FA9550-21-1-0340.
\section*{Author Declarations} \label{sec:Author Declarations}
\subsection*{Conflict of Interest}
The authors have no conflicts to disclose

\subsection*{Author Contributions}
\textbf{Ryan C. Baker:} Conceptualization (supporting); Data curation (equal); Formal analysis (equal); Investigation (lead); Methodology (equal); Project administration (equal); Software (lead); Validation (equal); Visualization (lead); Writing - original draft (lead); Writing - review \& editing (equal).
\textbf{Bridget O'Mara:} Investigation (supporting); Validation (supporting); Writing - review \& editing (supporting).
\textbf{Jacob L. Roberts:} Conceptualization (lead); Data curation (equal); Formal analysis (equal); Funding acquisition (lead); Methodology (equal); Project administration (equal); Resources (lead); Software (supporting); Supervision (lead); Validation (equal); Visualization (supporting); Writing - original draft (supporting); Writing - review \& editing (equal).
\section*{Data Availability} \label{sec:Data Availability}
The data that support the findings of this study are available from the corresponding author upon reasonable request.
\section*{References}
\bibliography{refs}

\end{document}